# Nanosecond nanothermometry in an electron microscope

*Florian Castioni[1], Yves Auad[1], Jean-Denis Blazit[1], Xiaoyan Li[1], Steffi Y. Woo [1,2], Kenji Watanabe[3], Takashi Taniguchi[4], Ching-Hwa Ho[5], Odile Stéphan[1], Mathieu Kociak[1], Luiz H.G. Tizei[1,\*]*

[1]Univ. Paris-Saclay, CNRS, Laboratoire de Physique des Solides, 91405, Orsay, France

[2]Center for Nanophase Materials Sciences, Oak Ridge National Laboratory, Oak Ridge, TN, 37381, U.S.A.

[3]Research Center for Electronic and Optical Materials, National Institute for Materials Science, 1-1 Namiki, Tsukuba 305-0044, Japan

[4]Research Center for Materials Nanoarchitectonics, National Institute for Materials Science, 1-1 Namiki, Tsukuba 305-0044, Japan

[5]Graduate Institute of Applied Science and Technology, National Taiwan University of Science and Technology, Taipei 106, Taiwan

ABSTRACT

Thermal transport in nanostructures plays a critical role in modern technologies. As devices shrink, techniques that can measure thermal properties at nanometer and nanosecond scales are increasingly needed to capture transient, out-of-equilibrium phenomena. We present a novel pump-probe photon-electron method within a scanning transmission electron microscope (STEM) to map temperature dynamics with unprecedented spatial and temporal resolutions. By combining focused laser-induced heating and synchronized time-resolved monochromated electron energy loss spectroscopy (EELS), we track phonon, exciton and plasmon signals in various materials, including silicon nitride, aluminum thin film, and transition metal dichalcogenides. Our results demonstrate the technique's ability to follow temperature changes at the nanometer and nanosecond scales. The experimental data closely matched theoretical heat diffusion models, confirming the method's validity. This approach opens new opportunities to investigate transient thermal phenomena in nanoscale materials, offering valuable insights for applications in thermoelectric devices and nanoelectronics.

In recent years, the advancement of nanoelectronics, photonics, and thermoelectric materials has intensified the need for precise thermal transport management at nanoscale dimensions. As electronic devices become increasingly miniaturized, efficient dissipation of heat is critical to ensure performance, stability, and longevity.[1] In nanoscale materials, the heat transport mechanisms deviate significantly from bulk behavior, as the geometries and reduced dimensions introduce boundary scattering and phonon confinement effects,[2] necessitating a deeper exploration of these phenomena.

To address these challenges, a broad range of spatially resolved thermal characterization techniques are existing. Optical techniques, including thermoreflectance[3] and Raman spectroscopy,[4] offer non-contact solutions with good sensitivity and potentially high temporal-resolution. However, the optical diffraction limit restricts their spatial resolution, often to around several hundred nanometers. Advancements in local probe techniques, and particularly scanning thermal microscopy (SThM)[5] have opened up new possibilities for localized temperature mapping within tens of nanometer spatial-resolution, although these methods can be challenged by contact artifacts and sensitivity constraints. Alternatively, electron microscopy-related techniques such as diffraction[6], electron energy loss spectroscopy (EELS)[7–10] or cathodoluminescence (CL)[11–13] offer valuable insights as the spatial resolution offered by fast electrons permit to map material temperature distribution with ultimately nanometer spatial resolution. A few studies have explored the potential of injecting light into the microscope to investigate the effects of temperature increases on spectroscopy results.[14–16] However, beyond static properties, there is a pressing need to investigate transient, out-of-equilibrium behavior on timescales ranging from picoseconds to microseconds. Current techniques able to map the temperature at both nanometer and sub-microsecond scales remain limited.

In this study, we introduce a novel pump-probe photon-electron method based on synchronized electron detection for measuring the temperature of materials with nanometer and nanosecond resolutions in a scanning transmission electron microscope (STEM). It relies on a unique combination of diffraction limited laser injection for triggering local temperature change, and an EELS experimental setup with a few tens of meV energy resolution, 1.5 ns temporal resolution, and sub nanometer spatial resolution for mapping thermal changes. The pump-probe is implemented by tracking the temporal delay between the detection of an electron having lost energy as detected by EELS and the previous laser pulse excitation. More specifically, a ~1 µm and ~25 ns visible-range pulsed laser is focused on an area probed with a 100 keV electron beam. The laser-induced heating and subsequent cooling process is studied in three different materials, utilizing distinct electron scattering mechanisms to estimate the local temperatures changes (**Figure 1a**). Phononic signal in the mid-IR in a silicon nitride slab is used to validate quantitatively the principle of the method as temperature changes can be directly measured from first principles (detailed balance). Then, the variations of the bulk plasmon resonance's energy are exploited, as they are related to temperatures changes. Measuring them in time and space (**Figure 1b**) in an aluminum slab permits to follow the heating dynamic with ns and nm resolutions. Finally, we explore the heating and cooling dynamics in a transition metal dichalcogenides (TMDs) monolayer by following the variation of the exciton energy values and linking them to temperature changes thanks to previous heating experimental data.

Furthermore, we apply a simple classical 2D heat diffusion model to successfully capture the thermal dynamics observed at nanometer and nanosecond scales, using laser pulse intensity as the only adjustable parameter to fit the experimental data. Excellent agreement is achieved for the thin aluminum foil sample (**Figure 1c**), whose simple geometry allows this 2D thermal diffusion model to accurately predict the heat dynamics of the experiment.

The experiment was conducted on a monochromated, aberration-corrected Nion Hermes 200 STEM at 100 kV, with an energy resolution of ~50 meV and ~10 pA beam current. Laser pulses (~25 ns with wavelength in the 580–610 nm range) were injected using a parabolic mirror. Time-resolved EELS spectra were recorded using a Timepix3 event-based camera, which enables electron detection with a temporal accuracy of approximately 1.5 ns, providing the temporal resolution for the experiment.[17,18] The details of the experimental set up are provided in the supplementary material (SI). In an EELS experiment, a fast electron beam transfers energy to the sample of interest, creating various excitations, such as phonons in the infra-red range, excitons in the visible range and bulk plasmons in the far-UV energy range. As exemplified in the **Figure 1**, an EELS spectrum consists of a zero-loss peak (ZLP), made up of elastically scattered electrons, loss peaks at energies characteristic of the various excitations, and optionally of a detectable gain peak in the infra-red related to thermally excited phonons, as explained later. Modern monochromated microscopes have two key features enabling the experiments presented here. First, monochromation reduces the intensity of the ZLP tails, allowing to access the infra-red regime. Second, the increased spectral resolution permits to resolve phonon and exciton linewidths. Finally, temporal resolution is achieved by tracking the time delay between the detection of an electron in EELS using an event-based detector and the pulsed laser excitation. We stress that in our experiment the electron source is not pulsed, but is a standard continuous cold field emission source. The samples were maintained at room temperature, except for the TMD sample, which was cooled to liquid nitrogen temperature (~100 K) to limit the damage caused by the laser.

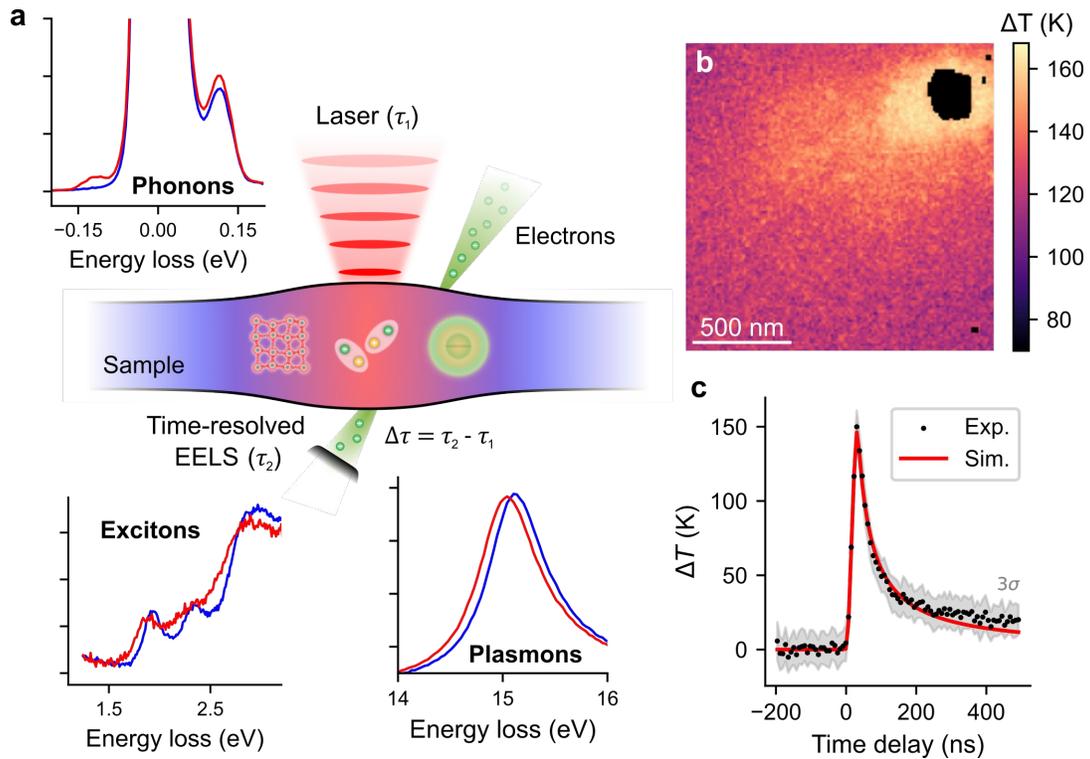

**Figure 1. Phonons, excitons and plasmons for nanosecond-resolved nanothermometry**. (a) A 25 ns width pulsed laser in the visible range heats the sample, inducing absorption changes probed by EELS in the IR (phonons) to the soft X-ray (bulk plasmon) energy range in different material systems. (b) Temperature difference spatial mapping of an aluminum thin film following laser exposure. (c) Temperature temporal profile (including 3σ uncertainty) of the aluminum film depending on the time delay $\Delta t$ following the laser excitation, compared with a heat diffusion simulation in an infinite 2D film.

We start by assessing that our methodology can follow quantitatively the temperature dynamics. We use the fact that, if one can access the infra-red regime, a material absolute temperature can be quantified from an EELS spectrum by using the principle of detailed balance (PDB).[8,9] This principle establishes that the ratio between spontaneous loss and thermal gain energy peaks are directly related to the material temperature T (the details of the PDB theory are provided in SI). We have used this methodology to quantify the temperature of a holey amorphous $SiN_x$ membrane using the previously described pump-probe scheme. The electron beam probes the phonon signal evolution over time in an aloof configuration, positioned outside the material. In the case of the $SiN_x$ membrane, the electron beam interacts predominantly with the longitudinal optical phonon of the material around ~ 118 meV, in excellent agreement with previously reported values from density function theory (DFT).[19]

Before the laser pulse, the EELS spectrum mainly exhibits a peak related to the phonon on the energy loss side (**Figure 2a**). Following the PDB, both phonon energy gain and loss peaks increased in intensity just after the laser excitation, and progressively decrease hundreds of nanoseconds after the pulse. Additionally, the ZLP appears to broaden, likely due to higher occupancy of lower-energy phonon states. By tracking the evolution of the peak ratio over time, the measured temperature increases up to $T_{max} = 680 \pm 40\ K$ at the end of the 25 ns pulse (**Figure 2b**). A rapid cooling phase follows during the first few hundred nanoseconds, transitioning into a slower cooling regime over several microseconds.

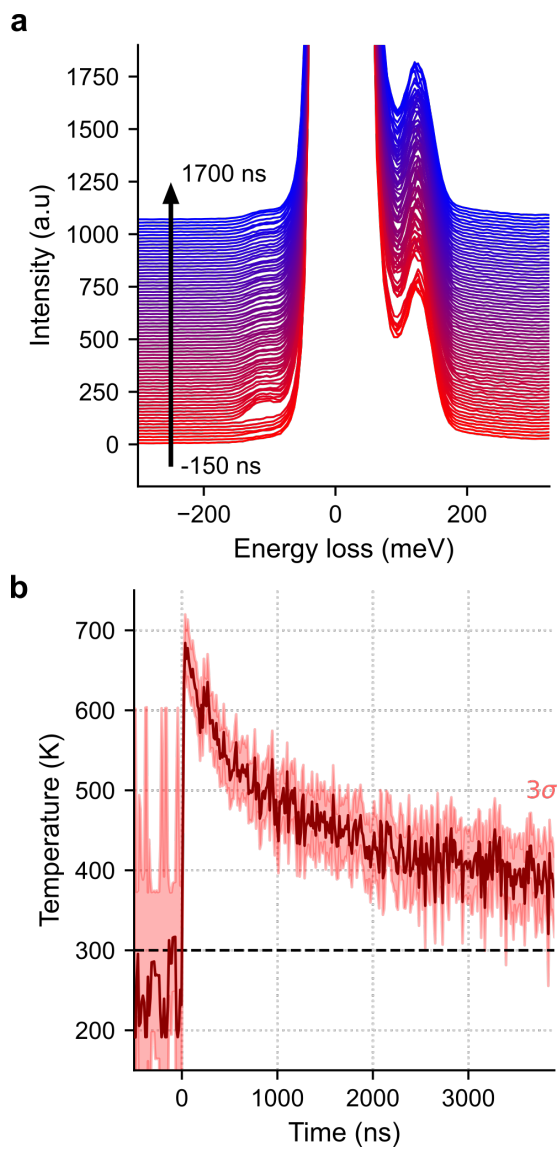

**Figure 2. Absolute temperature measurement in a SiN$_x$ thin film derived from the principle of detailed balance (PDB).** (a) Low-energy EELS cascade spectra showing the evolution of SiN$_x$ optical phonon with time, where $t = 0$ s corresponds to the onset of the laser pulse. A significant change is observed in both the spontaneous energy loss and gain peaks. (b) Absolute temperature temporal profile (including 3σ uncertainty) extracted from PDB, illustrating the cooling dynamics of the material following the photon pulse, measured over a long timescale (~μs).

Having validated our methodology, we extend it to the measurement of spatio-temporal temperature changes with nm and ns resolutions. With this aim, we use bulk plasmon energy shifts as they have long been recognized as reliable spectroscopic markers for indirectly measuring the temperature of semiconductors and metals using EELS,[20,21] a technique sometimes referred to as plasmon energy expansion thermometry (PEET).[7] The method relies on the shift in plasmon resonance energy caused by the thermal expansion of the material, which affects the density of the electron gas. Metals - particularly aluminum - are excellent candidates for such experiments, as their high density of conduction band electrons can be accurately modeled using a free electron model, and they exhibit relatively sharp plasmon resonances.

This approach is used to resolve both temporally and spatially the temperature evolution of an aluminum thin film (~20 nm of thickness, deposited by evaporation on a few nanometers thick graphite supported over a conventional TEM grid), following the laser excitation. The bulk plasmon resonance, located at 15.22 eV (in excellent agreement with previously reported values[21,22]), exhibits a red-shift of ~0.08 eV after the laser pulse (**Figure 3a**). The peak then recovers progressively towards the pre-excitation energy value, witnessing the quick diffusion of the heat in the metallic film.

The relationship between this bulk plasmon energy shift and the material's temperature variation can be made using the Drude model to describe the collective resonances of free electrons. The change in free electrons density can be associated with the temperature through the thermal expansion coefficient's evolution with temperature (detailed theory is provided in SI). Because of the higher signal intensity (compared to the phonon peaks studied earlier), we achieve a much better temporal resolution, enabling us to clearly resolve both the heating and cooling phases of the transition. The maximal temperature rise ($\Delta T = 150 \pm 10$ K) is reached at the end of the laser pulse, followed by rapid heat dissipation in the first tens of nanoseconds (**Figure 1c**). The cooling process then continues over hundreds of seconds, until reaching the pre-heating equilibrium state. These observations are well supported by a simple heat diffusion model based on Fourier equation solved inside a 2D infinite aluminum film (**Figure 1c**). In the simulations, only the laser power input is kept as a free parameter to match the maximum temperature rise at the end of the pulse. Then, the cooling process is only dependent on the material parameters, especially its thermal conductivity. The excellent agreement between experimental measurements and simulation confirms that our methodology permits to precisely follow temperature variations in space and time and could be used to tackle more complicated geometries.

A key aspect of our setup relies on the focused laser beam, which produces a spot smaller than the typical field of view in STEM. Hence, a gradient of temperature imprinted on the metallic film by the laser beam profile is expected and can be measured. To this purpose, spatially resolved time-energy histogram, similar to **Figure 3a,** are recorded while scanning a 2 x 2 µm² area around the laser central spot. The heating process can then be observed through the spatially resolved temporal temperature profiles (identical to **Figure 1c**) extracted at each electron beam position.

The first panel in **Figure 3b** shows the high-angle annular dark field (HAADF) image of the scan area, where a ~100 nm hole is visible due to prior laser irradiation. The bright contrast around the hole is a consequence of partial recrystallisation of the film caused by the heat. As expected, the area near the hole experiences the strongest heating, with temperature decreasing further away (**Figure 1b**). The other panels in **Figure 3b** show the evolution of $\Delta T$ over time. Before the pulse ($t = -4$ ns), the film presents a homogeneous temperature. During the pulse (within the first 25 ns), significant temperature inhomogeneity arises due to the photon beam shape. The temperature profiles extracted along the white arrow in **Figure 3b** indicates that a temperature gradient of ~50 K over 1.5 µm is created (first panel of **Figure 3c**). After the photon pulse, approximately 15 ns (frame t = 42 ns) is sufficient for the film to exhibit a complete temperature homogenization over the scanned area. The whole area then continues to cool down by heat conduction with the surrounding (i.e. outside of scanned) areas. After 400 ns, the material has almost returned to room temperature.

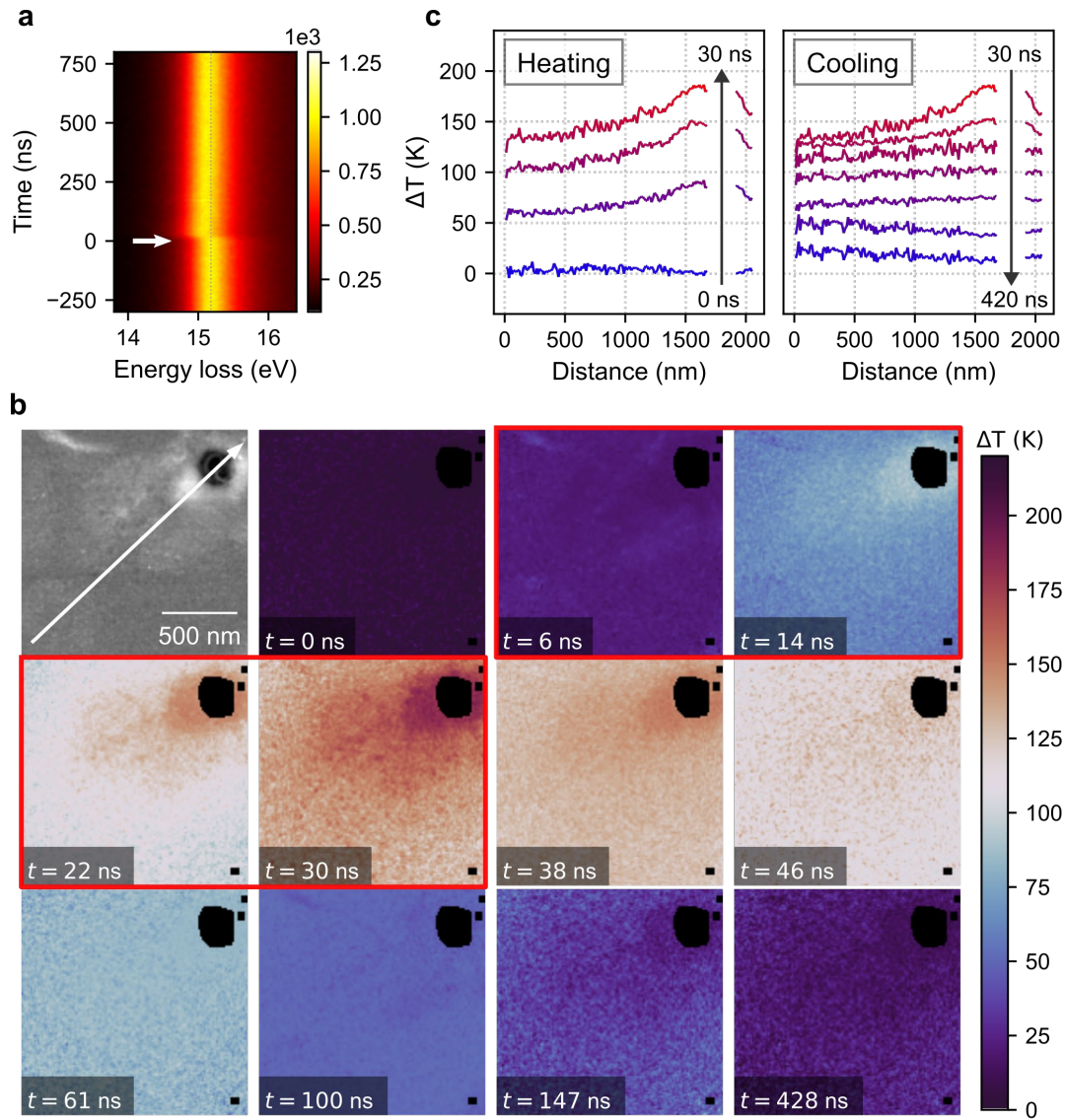

**Figure 3. Temporal and spatial temperature variations in a thin Al film excited by a 25 ns laser pulse.** (a) Time-energy histogram of the aluminum bulk plasmon following the photon pulse (white arrow), showing a pronounced peak shift due to local heating effects. The dotted line marks the plasmon energy before the pulse arrival, providing a clear reference for visualizing the shift. (b) HAADF (top left) and differential temperature maps of the film at various time delays after the pulse. The first panel corresponds to the dark field image, acquired simultaneously. The red framed maps are acquired during the laser pulse. (c) Spatial distribution of the temperature rise along the linescan - indicated by the white arrow in (b) - as a function of time. The two panels depict the heating phase (pulse on) and the cooling phase (pulse off), respectively.

The method can be used to study the dynamic thermal response of lesser-known materials. We measured the dynamic response after photoexcitation in a monolayer of TMD, a direct bandgap semiconductor with promising properties for the development of next-generation reduced-dimensional optical devices.[23] Assessing nanoscale heating properties in these materials is highly relevant. These 2D semiconductors exhibit strong excitonic properties, which have been intensively studied via EELS.[24] The sample consists of a monolayer of $WS_2$, encapsulated between two thin (~10 nm) hBN flakes, used to protect the sample and homogenize the dielectric environment of the semiconductor. The layers were assembled using the standard dry-stamping transfer technique, and the full stack was deposited in a planar configuration on a carbon quantifoil TEM grid (sample images are available in the SI).[25]

The pump-probe photon-electron experiment was performed on this $hBN/WS_2/hBN$ heterostructure. Before the pulse, the EELS spectrum presents 3 different absorption peaks in the visible range between 1 and 4 eV (**Figure 4a**), related to the principal exciton resonances (referred to as A, B, and C) located at 2.01 eV, 2.40 eV and 2.89 eV, respectively. The first two peaks are associated with direct transitions through the band gap in the K and K' points, and present different energies related to the strong spin–orbit coupling characteristic of these materials. Exciton C reflects contributions from multiple direct transitions away from K point, including between Γ and Q points,[26] near Q point and between K and M points.[27]

After the photon pulse at t = 0 s, all three peaks shift toward lower energies (**Figure 4a**). The B and C exciton peaks are completely broadened, while the A exciton remains discernible from the background. These observations are clearer in the cascade EELS spectra (**Figure 4b**), where each spectrum corresponds to a 50 ns time window integration to improve the signal-to-noise ratio. The second spectrum after the pulse clearly illustrates the disappearance of the B exciton peak. Initially, the A exciton appears unchanged, but shifts to lower energies only after the end of the pulse. This effect is attributed to the appearance of a stimulated energy loss signal during the pulse, related to interaction between the electron beam and the evanescent electromagnetic field generated by the laser on the material. The observation of this effect in a semiconductor, called photon induced near-field electron microscopy (PINEM), remains unclear and requires further investigation.

As in most of the common semiconductors, the temperature increase leads to a reduction of its electronic bandgap, a direct consequence of the thermal expansion experienced by the material. The temperature effect on the optical signature of TMDs has been previously studied through optical absorption and photoluminescence spectroscopies,[28–30] demonstrating indeed a significant A exciton energy shift towards the lower energies when the temperature increase. Tizei et al.[31] and Gogoi *et al.*[32] observed a similar trend in EELS of freestanding TMD monolayers, and also reported the disappearance of the B exciton with increasing temperature as observed in the current study. A linear redshift of the exciton resonances was reported, with a consistent coefficient of approximately 0.5 meV/K for all the types of TMDs studied.[31,32] In the current study, this empirical measurement is applied to translate the measured A exciton shift into a temperature rise. The peak was fitted using a Gaussian function, and its center evolution over time – and the temperature rising herewith measured – is plotted in **Figure 4c**. A significant temperature increase of more than $\Delta T = 220 \pm 50$ K was observed. The material's temperature then dropped by 100 K after ~500 ns following the laser heating, followed by continued heat dissipation over a longer time scale. In this heterostructure, predicting the temperature dynamics is particularly challenging due to the diverse nature of the materials involved. Since hBN has a bandgap of approximately 6 eV, significant part of the light absorption is expected to occur within the direct band-gap $WS_2$ monolayer. However, with a thermal conductivity more than one order of magnitude higher than that of WS$_2$,[33,34] hBN is expected to rapidly dissipate heat across its layers.

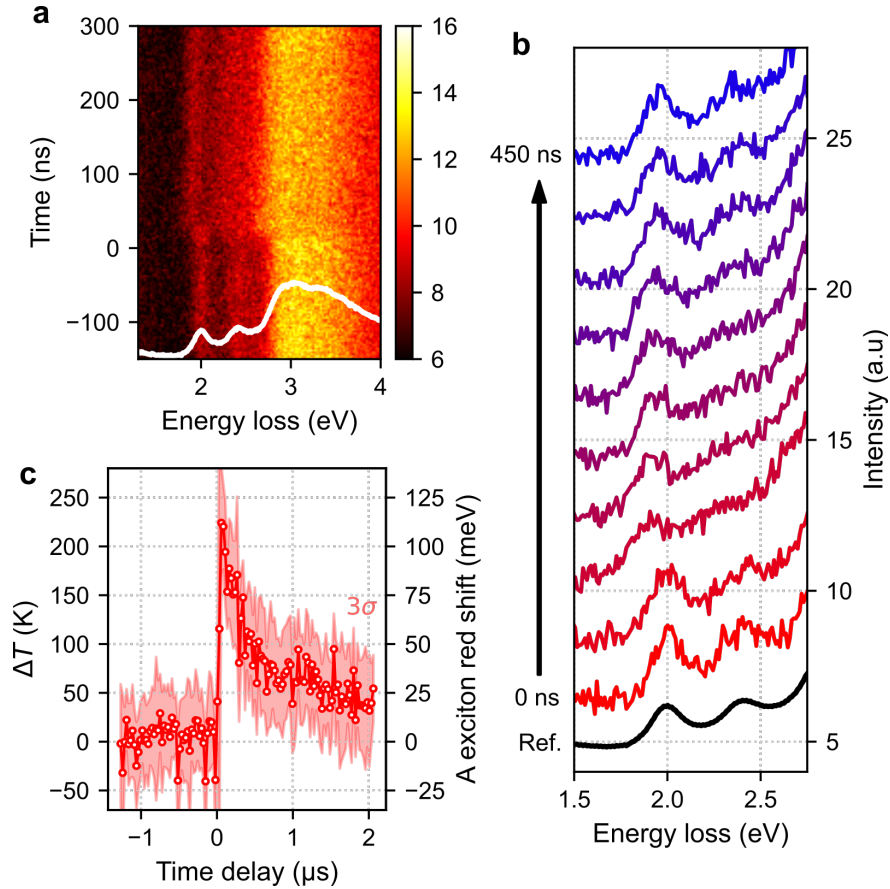

**Figure 4. Exciton peak shift following a 25 ns laser excitation.** (a) Time-energy histogram of WS$_2$ excitons after photon pulse. The integrated white spectrum corresponds to the regular EELS spectrum before excitation. (b) Cascade EELS spectra as a function of time, showing the evolution of excitons A and B during the heating and cooling processes. Each spectrum corresponds to an integration window of 50 ns. (c) Temperature temporal profile (including 3σ uncertainty) extracted from the A exciton red shift measurements.

In conclusion, we have demonstrated a novel pump-probe approach combining photons and electrons to measure temperature dynamics in materials with nanosecond time resolution and nanometer spatial resolution. The method combines diffraction-limited light injection, highly monochromatized and event-based EELS. Applied to three distinct materials, this technique revealed significant differences in thermal responses, despite similar laser power density inputs. Each material was probed using distinct excitation processes spanning different energies with electron energy-loss spectroscopy, showing the possible applications to many different systems. A 2D heat conduction simulation showed to be in particularly good agreement with the temperature measurements performed with nm and ns resolutions. This validates the principle of our method that should be applied to more complex materials and nanodevices in the future. Thanks to the excellent spatial resolution of TEM, this technique enables the correlation of intrinsic material properties - such as interfaces, defects, and chemical inhomogeneities - with thermal properties. This technique offers a means to probe anisotropic heat transport down to the nanometer scale. Notably, this capability allows for the investigation of how heat diffusion is influenced by material geometry, as recently demonstrated in studies of asymmetric nanoparticle assemblies, where nanoparticle asymmetric geometry induced a symmetry break in heat transport even under isotropic heating.[35]

To further optimize the accuracy of the technique, several experimental aspects need refinement. To accurately quantify potential measurement errors, comparisons with *in situ* stimuli - using either a heating holder or nanodevice biasing[7] - will be essential. Calibrating the laser power input carefully is equally crucial to avoid damaging thin films and to better understand the heating dynamics. Additionally, the mechanisms of photon absorption driving temperature changes in these materials require further investigation, which could be achieved by tuning the laser wavelength, an option already available on our current setup.[36] Looking ahead, this technique offers the potential for even finer time resolution, down to the picosecond scale, while still capturing critical spectroscopic signals, as allowed by the Heisenberg uncertainty principle (1 ps.meV).

ASSOCIATED CONTENT

AUTHOR INFORMATION

**Corresponding Author**

Corresponding author: luiz.tizei@cnrs.fr (L. H. G. Tizei)

**Author Contributions**

The manuscript was written through contributions of all authors. All authors have given approval to the final version of the manuscript.

SUPPORTING INFORMATION

Additional experimental details, theoretical descriptions for PDB and PEET, heat diffusion model, and data treatment methodology.

ACKNOWLEDGMENT

This project has been funded in part by the European Union through the Horizon 2020 Research and Innovation Program (grant agreement No. 101017720 (EBEAM)) and by the French National Agency for Research under the program of future investment TEMPOS-CHROMATEM (reference no. ANR-10-EQPX-50) and the JCJC grant SpinE (reference no. ANR-20-CE42-0020). K.W. and T.T. acknowledge support from the JSPS KAKENHI (Grant Numbers 21H05233 and 23H02052) and World Premier International Research Center Initiative (WPI), MEXT, Japan.

ABBREVIATIONS

(HA)ADF, (high-angle) annular dark field; CL, cathodoluminescence; DFT, density functional theory; EELS, electron energy-loss spectroscopy; PDB, principle of detailed balance; PEET, plasmon energy expansion thermometry; PINEM, photon induced near-field electron microscopy; STEM, scanning transmission electron microscopy; SThM, scanning thermal microscopy; TMD, transition metal dichalcogenides; ZLP, zero-loss peak.

REFERENCES

(1) Tong, X. C. *Advanced Materials for Thermal Management of Electronic Packaging*; Springer Series in Advanced Microelectronics; Springer New York: New York, NY, 2011; Vol. 30. https://doi.org/10.1007/978-1-4419-7759-5.

(2) El Sachat, A.; Alzina, F.; Sotomayor Torres, C. M.; Chavez-Angel, E. Heat Transport Control and Thermal Characterization of Low-Dimensional Materials: A Review. *Nanomaterials* **2021**, *11* (1), 175. https://doi.org/10.3390/nano11010175.

(3) Rosencwaig, A.; Opsal, J.; Smith, W. L.; Willenborg, D. L. Detection of Thermal Waves through Optical Reflectance. *Applied Physics Letters* **1985**, *46* (11), 1013–1015. https://doi.org/10.1063/1.95794.

(4) Malekpour, H.; Balandin, A. A. Raman-based Technique for Measuring Thermal Conductivity of Graphene and Related Materials. *J Raman Spectroscopy* **2018**, *49* (1), 106–120. https://doi.org/10.1002/jrs.5230.

(5) Menges, F.; Mensch, P.; Schmid, H.; Riel, H.; Stemmer, A.; Gotsmann, B. Temperature Mapping of Operating Nanoscale Devices by Scanning Probe Thermometry. *Nat Commun* **2016**, *7* (1), 10874. https://doi.org/10.1038/ncomms10874.

(6) Niekiel, F.; Kraschewski, S. M.; Müller, J.; Butz, B.; Spiecker, E. Local Temperature Measurement in TEM by Parallel Beam Electron Diffraction. *Ultramicroscopy* **2017**, *176*, 161–169. https://doi.org/10.1016/j.ultramic.2016.11.028.

(7) Mecklenburg, M.; Hubbard, W. A.; White, E. R.; Dhall, R.; Cronin, S. B.; Aloni, S.; Regan, B. C. Nanoscale Temperature Mapping in Operating Microelectronic Devices. *Science* **2015**, *347* (6222), 629–632. https://doi.org/10.1126/science.aaa2433.

(8)  Idrobo, J. C.; Lupini, A. R.; Feng, T.; Unocic, R. R.; Walden, F. S.; Gardiner, D. S.; Lovejoy, T. C.; Dellby, N.; Pantelides, S. T.; Krivanek, O. L. Temperature Measurement by a Nanoscale Electron Probe Using Energy Gain and Loss Spectroscopy. *Phys. Rev. Lett.* **2018**, *120* (9), 095901. https://doi.org/10.1103/PhysRevLett.120.095901.

(9)  Lagos, M. J.; Batson, P. E. Thermometry with Subnanometer Resolution in the Electron Microscope Using the Principle of Detailed Balancing. *Nano Lett.* **2018**, *18* (7), 4556–4563. https://doi.org/10.1021/acs.nanolett.8b01791.

(10) Hu, X.; Yasaei, P.; Jokisaari, J.; Öğüt, S.; Salehi-Khojin, A.; Klie, R. F. Mapping Thermal Expansion Coefficients in Freestanding 2D Materials at the Nanometer Scale. *Phys. Rev. Lett.* **2018**, *120* (5), 055902. https://doi.org/10.1103/PhysRevLett.120.055902.

(11) Mauser, K. W.; Solà-Garcia, M.; Liebtrau, M.; Damilano, B.; Coulon, P.-M.; Vézian, S.; Shields, P. A.; Meuret, S.; Polman, A. Employing Cathodoluminescence for Nanothermometry and Thermal Transport Measurements in Semiconductor Nanowires. *ACS Nano* **2021**, *15* (7), 11385–11395. https://doi.org/10.1021/acsnano.1c00850.

(12) Park, W.-W.; Olshin, P. K.; Kim, Y.-J.; Nho, H.-W.; Mamonova, D. V.; Kolesnikov, I. E.; Medvedev, V. A.; Kwon, O.-H. Nanoscale Cathodoluminescence Thermometry with a Lanthanide-Doped Heavy-Metal Oxide in Transmission Electron Microscopy. *ACS Nano* **2024**, *18* (6), 4911–4921. https://doi.org/10.1021/acsnano.3c10020.

(13) Sáenz De Santa María Modroño, P.; Girard, H. A.; Arnault, J.; Jacopin, G. High-Resolution Thermal Sensing Using Temperature-Sensitive Cathodoluminescence Spectroscopy in Nitrogen-Doped Nanodiamonds. *Physica Status Solidi (a)* **2024**, 2400573. https://doi.org/10.1002/pssa.202400573.

(14)   Carbone, F.; Kwon, O.-H.; Zewail, A. H. Dynamics of Chemical Bonding Mapped by Energy-Resolved 4D Electron Microscopy. *Science* **2009**, *325* (5937), 181–184. https://doi.org/10.1126/science.1175005.

(15)   Rossouw, D.; Bugnet, M.; Botton, G. A. Structural and Electronic Distortions in Individual Carbon Nanotubes under Laser Irradiation in the Electron Microscope. *Phys. Rev. B* **2013**, *87* (12), 125403. https://doi.org/10.1103/PhysRevB.87.125403.

(16)   Van Der Veen, R. M.; Penfold, T. J.; Zewail, A. H. Ultrafast Core-Loss Spectroscopy in Four-Dimensional Electron Microscopy. *Structural Dynamics* **2015**, *2* (2), 024302. https://doi.org/10.1063/1.4916897.

(17)   Auad, Y.; Walls, M.; Blazit, J.-D.; Stéphan, O.; Tizei, L. H. G.; Kociak, M.; De la Peña, F.; Tencé, M. Event-Based Hyperspectral EELS: Towards Nanosecond Temporal Resolution. *Ultramicroscopy* **2022**, *239*, 113539. https://doi.org/10.1016/j.ultramic.2022.113539.

(18)   Auad, Y.; Baaboura, J.; Blazit, J.-D.; Tencé, M.; Stéphan, O.; Kociak, M.; Tizei, L. H. G. Time Calibration Studies for the Timepix3 Hybrid Pixel Detector in Electron Microscopy. *Ultramicroscopy* **2024**, *257*, 113889. https://doi.org/10.1016/j.ultramic.2023.113889.

(19)   Legut, D.; Wdowik, U. D.; Kurtyka, P. Vibrational and Dielectric Properties of α-Si3N4 from Density Functional Theory. *Materials Chemistry and Physics* **2014**, *147* (1–2), 42–49. https://doi.org/10.1016/j.matchemphys.2014.03.058.

(20)   Rocca, M.; Moresco, F.; Valbusa, U. Temperature Dependence of Surface Plasmons on Ag(001). *Phys. Rev. B* **1992**, *45* (3), 1399–1402. https://doi.org/10.1103/PhysRevB.45.1399.

(21)   Abe, H.; Terauchi, M. Temperature Dependence of the Volume-Plasmon Energy in Aluminum. *Journal of Electron Microscopy* **1992**. https://doi.org/10.1093/oxfordjournals.jmicro.a050994.

(22) Chmielewski, A.; Ricolleau, C.; Alloyeau, D.; Wang, G.; Nelayah, J. Nanoscale Temperature Measurement during Temperature Controlled in Situ TEM Using Al Plasmon Nanothermometry. *Ultramicroscopy* **2020**, *209*, 112881. https://doi.org/10.1016/j.ultramic.2019.112881.

(23) Huang, L.; Krasnok, A.; Alú, A.; Yu, Y.; Neshev, D.; Miroshnichenko, A. E. Enhanced Light–Matter Interaction in Two-Dimensional Transition Metal Dichalcogenides. *Rep. Prog. Phys.* **2022**, *85* (4), 046401. https://doi.org/10.1088/1361-6633/ac45f9.

(24) Woo, S. Y.; Galvao Tizei, L. H. Nano-Optics of Transition Metal Dichalcogenides and Their van Der Waals Heterostructures with Electron Spectroscopies. *2D Mater.* **2024**. https://doi.org/10.1088/2053-1583/ad97c8.

(25) Shao, F.; Woo, S. Y.; Wu, N.; Schneider, R.; Mayne, A. J.; De Vasconcellos, S. M.; Arora, A.; Carey, B. J.; Preuß, J. A.; Bonnet, N.; Och, M.; Mattevi, C.; Watanabe, K.; Taniguchi, T.; Niu, Z.; Bratschitsch, R.; Tizei, L. H. G. Substrate Influence on Transition Metal Dichalcogenide Monolayer Exciton Absorption Linewidth Broadening. *Phys. Rev. Materials* **2022**, *6* (7), 074005. https://doi.org/10.1103/PhysRevMaterials.6.074005.

(26) Carvalho, A.; Ribeiro, R. M.; Castro Neto, A. H. Band Nesting and the Optical Response of Two-Dimensional Semiconducting Transition Metal Dichalcogenides. *Phys. Rev. B* **2013**, *88* (11), 115205. https://doi.org/10.1103/PhysRevB.88.115205.

(27) Hong, J.; Koshino, M.; Senga, R.; Pichler, T.; Xu, H.; Suenaga, K. Deciphering the Intense Postgap Absorptions of Monolayer Transition Metal Dichalcogenides. *ACS Nano* **2021**, *15* (4), 7783–7789. https://doi.org/10.1021/acsnano.1c01868.

(28) Tongay, S.; Zhou, J.; Ataca, C.; Lo, K.; Matthews, T. S.; Li, J.; Grossman, J. C.; Wu, J. Thermally Driven Crossover from Indirect toward Direct Bandgap in 2D Semiconductors:

MoSe $_2$ versus MoS $_2$. *Nano Lett.* **2012**, *12* (11), 5576–5580. https://doi.org/10.1021/nl302584w.

(29)     Zhao, W.; Ribeiro, R. M.; Toh, M.; Carvalho, A.; Kloc, C.; Castro Neto, A. H.; Eda, G. Origin of Indirect Optical Transitions in Few-Layer MoS $_2$ , WS $_2$ , and WSe $_2$. *Nano Lett.* **2013**, *13* (11), 5627–5634. https://doi.org/10.1021/nl403270k.

(30)     Arora, A.; Wessling, N. K.; Deilmann, T.; Reichenauer, T.; Steeger, P.; Kossacki, P.; Potemski, M.; Michaelis De Vasconcellos, S.; Rohlfing, M.; Bratschitsch, R. Dark Trions Govern the Temperature-Dependent Optical Absorption and Emission of Doped Atomically Thin Semiconductors. *Phys. Rev. B* **2020**, *101* (24), 241413. https://doi.org/10.1103/PhysRevB.101.241413.

(31)     Tizei, L. H. G.; Lin, Y.-C.; Lu, A.-Y.; Li, L.-J.; Suenaga, K. Electron Energy Loss Spectroscopy of Excitons in Two-Dimensional-Semiconductors as a Function of Temperature. *Applied Physics Letters* **2016**, *108* (16), 163107. https://doi.org/10.1063/1.4947058.

(32)     Gogoi, P. K.; Lin, Y.-C.; Senga, R.; Komsa, H.-P.; Wong, S. L.; Chi, D.; Krasheninnikov, A. V.; Li, L.-J.; Breese, M. B. H.; Pennycook, S. J.; Wee, A. T. S.; Suenaga, K. Layer Rotation-Angle-Dependent Excitonic Absorption in van Der Waals Heterostructures Revealed by Electron Energy Loss Spectroscopy. *ACS Nano* **2019**, *13* (8), 9541–9550. https://doi.org/10.1021/acsnano.9b04530.

(33)     Jiang, P.; Qian, X.; Yang, R.; Lindsay, L. **Anisotropic Thermal Transport in Bulk Hexagonal Boron Nitride**. *Phys. Rev. Materials* **2018**, *2* (6), 064005. https://doi.org/10.1103/PhysRevMaterials.2.064005.

(34) Zhang, Y.; Lv, Q.; Fan, A.; Yu, L.; Wang, H.; Ma, W.; Lv, R.; Zhang, X. Reduction in Thermal Conductivity of Monolayer WS2 Caused by Substrate Effect. *Nano Res.* **2022**, *15* (10), 9578–9587. https://doi.org/10.1007/s12274-022-4560-7.

(35) Feldman, M.; Vernier, C.; Nag, R.; Barrios-Capuchino, J. J.; Royer, S.; Cruguel, H.; Lacaze, E.; Lhuillier, E.; Fournier, D.; Schulz, F.; Hamon, C.; Portalès, H.; Utterback, J. K. Anisotropic Thermal Transport in Tunable Self-Assembled Nanocrystal Supercrystals. *ACS Nano* **2024**, *18* (50), 34341–34352. https://doi.org/10.1021/acsnano.4c12991.

(36) Auad, Y.; Dias, E. J. C.; Tencé, M.; Blazit, J.-D.; Li, X.; Zagonel, L. F.; Stéphan, O.; Tizei, L. H. G.; García De Abajo, F. J.; Kociak, M. μeV Electron Spectromicroscopy Using Free-Space Light. *Nat Commun* **2023**, *14* (1), 4442. https://doi.org/10.1038/s41467-023-39979-0.

# SUPPLEMENTARY MATERIAL

# Nanosecond nanothermometry in an electron microscope

*Florian Castioni[1], Yves Auad[1], Jean-Denis Blazit[1], Xiaoyan Li[1], Steffi Y. Woo [1,2], Kenji Watanabe[3], Takashi Taniguchi[4], Ching-Hwa Ho[5], Odile Stéphan[1], Mathieu Kociak[1], Luiz H.G. Tizei[1,*]*

[1]Univ. Paris-Saclay, CNRS, Laboratoire de Physique des Solides, 91405, Orsay, France

[2]Center for Nanophase Materials Sciences, Oak Ridge National Laboratory, Oak Ridge, TN, 37381, U.S.A.

[3]Research Center for Electronic and Optical Materials, National Institute for Materials Science, 1-1 Namiki, Tsukuba 305-0044, Japan

[4]Research Center for Materials Nanoarchitectonics, National Institute for Materials Science, 1-1 Namiki, Tsukuba 305-0044, Japan

[5]Graduate Institute of Applied Science and Technology, National Taiwan University of Science and Technology, Taipei 106, Taiwan

# Experimental setup

A schematic of the experiment is presented in **Figure 1a** of the article. The experiment was carried out in an aberration-corrected and monochromated Nion Hermes 200 scanning transmission electron microscope (STEM), operating at 100 kV. Different configurations of electron optics and EELS spectrometer setups were employed to achieve the spectral resolution required, as summarized in **Table S1**. Overall, the electron beam achieves a typical size of a few angstroms, significantly smaller than the scan step (~1.5 nm) used for mapping the heating of the aluminum film. The samples were maintained at room temperature, except for the TMDs stack, which was cooled to liquid nitrogen temperature (~100 K) to minimize laser-induced damage.

| Experiment | Convergence semi-angle (mrad) | EELS aperture (mm) | EELS dispersion (meV/pix) |
|---|---|---|---|
| Phonon on $SiN_x$ | 15 | 1 | 5 |
| Exciton on $WS_2$ | 25 | 2 | 10 |
| Plasmon on Al | 15 | 1 | 30 |

**Table S1**. Summary of microscope configurations for the three different experiments.

The sample is heated with a ~25 ns-pulsed dye laser in the wavelength range of 580–610 nm. The laser light was injected using an Attolight Mönch parabolic mirror with high (~0.5) numerical aperture, leading to highly astigmatic µm-sized spot. To align the electron beam and the light beam with the sample region of interest, the mirror was positioned using sub-micrometric precision motors and absolute encoders. The laser input power ranged between 20 and 60 µW, resulting in an average laser power density of ~$10^7$ W/m². The continuous electron beam was monochromated to an energy resolution of 50 meV, with a beam current of ~10 pA.

To observe the heat dynamic, the EELS spectra were detected in synchronization with the laser pulses using the CheeTah Timepix3 event-based detector, commercialized by Amsterdam Scientific Instruments. The latter also contains two time-to-digital converters (TDCs), used to timestamp external logic signals using the same reference clock used for timestamping incident electrons. This configuration allows for direct comparison between the electron detection time and external signals, particularly in our case laser pulses. The data were reconstructed with a search algorithm to sort the different detected events. Each electron can then be tagged by its energy loss ($\Delta E$), time delay relative to the laser pulse ($\Delta t$) and the corresponding spatial coordinates of the electron beam ($x$ and $y$). With a proper calibration of the camera, the maximal temporal resolution achieved was as short as ~1.5 ns.[1] Consequently, the acquisition speed of EELS hyperspectral images is not limited by the camera anymore, but rather by the scan coils. For the aluminum film mapping experiment, the typical dwell time used was 4 µs.

To accurately determine the pulse starting time compared to the electrons detection, we made use of the photon-induced near-field electron microscopy (PINEM) effect, where electrons get accelerated (and then gain energy) after interacting with the evanescent electromagnetic field of the material generated by the photon beam (**Figure S1b**). The evolution in time of the gain energy peak can then be used to directly access the time profile of the laser pulse (**Figure S1a**). Furthermore, the stimulated energy gain and loss allow for precise measurement of the actual energy dispersion of the magnetic prism, as their energies are directly related to the laser wavelength.[2,3] Assuming negligible non-linearity effects in the prism, we commonly observe a relative deviation of approximately 5% compared to the nominal value.

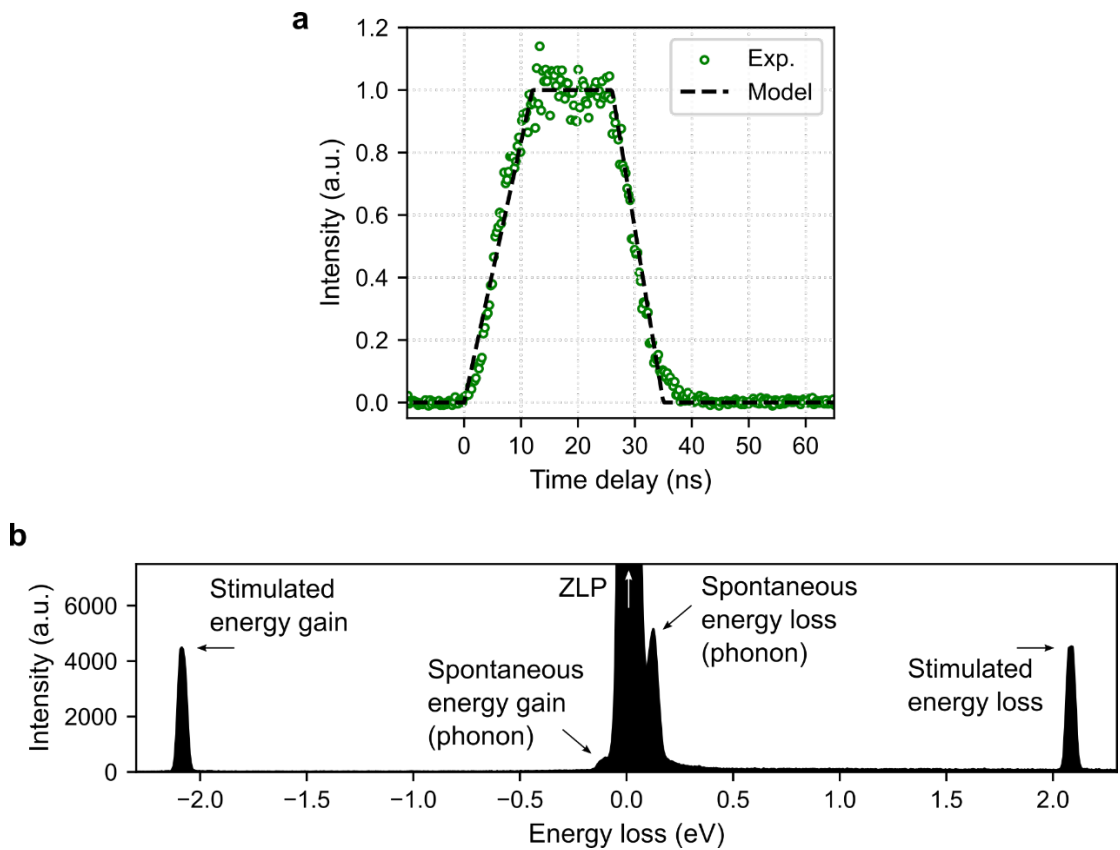

**Figure S1. Photon-induced near field electron microscopy (PINEM) effect on Si₃N₄ for time and energy calibration.** (a) Stimulated gain peak time profile serving as a template for the superimposed theoretical model (black dashed line) used as input for 2D heat conduction simulation. b) Electron energy loss spectrum (EELS) acquired during laser exposition, comparing stimulated (laser induced) and spontaneous (phonon scattering) gain and loss of energy.

# Theory of principle of detailed balance

The principle of detailed balance (PDB) establishes that, at temperature T at equilibrium, each elementary process is balanced by its reverse process. In EELS, it was shown that, based on statistical principles, PDB can be applied to the low-energy range to directly determine the absolute temperature by comparing the intensities of the spontaneous loss and gain peaks of a specific excitation $\Delta E$. If we note $P(\Delta E)$ and $P(-\Delta E)$ the energy loss and energy gain probabilities, the PDB is expressed as:

$$\frac{P(\Delta E)}{P(-\Delta E)} = e^{\Delta E/K_b T} \qquad (1)$$

where $K_b$ is the Boltzmann constant, and T is the absolute temperature in Kelvin. While PDB also applies to the momentum $\Delta q$ transferred to the material, this aspect is irrelevant in our case, as a circular aperture is placed before the spectrometer. The goal of this study is to resolve the time evolution of these signals after photoexcitation, enabling us to observe the heating and cooling processes in the material.

# PEET theory

According to the Drude model, the bulk plasmon energy $E_p$ is given by:[4]

$$E_p(T) = \hbar\omega_p(T) = \hbar\sqrt{\frac{4\pi\, n(T)\, e^2}{m}} \qquad (2)$$

where $\hbar$ is the reduced Planck constant, $\omega_p$ is the plasmon angular frequency, and $n$ is the temperature-dependent density of free electrons of charge $e$ and mass $m$. The variation in $n$ can be directly related to the material's linear thermal expansion coefficient $\alpha$. This coefficient is approximated with a linear dependence on temperature $\alpha(T) \approx \alpha_0 + \alpha_1(T - T_0)$, where $T_0$ is a reference temperature (conventionally RT) and $\alpha_l$ are parameters measured by Chmielewski *et al.*[5] over a temperature between 300 and 900 K. Note that a quadratic expression is usually used (especially in Ref. 5), but the linear approximation was adopted here to simplify the determination of temperature uncertainty. We verified that this simplification had minimal impact on the results within the considered temperature range. The relation between $n$ and T can then be expressed as:

$$n(T) \approx n(T_0)\left(1 - 3\int_{T_0}^{T}\alpha(T)dT\right) \qquad (3)$$

By combining equation (2) and (3), first-order approximation for the relation between $E_p$ and $T$ can be derived. To minimize uncertainty in the RT plasmon energy $E_p(T_0)$, we focus only on the relative change in temperature $\Delta T$ with respect to the relative plasmon shift $\Delta E_p$ before and after the photon pulse:

$$\Delta E_p(T) \approx -\frac{3}{2}\sum_{l=0}^{1}\frac{\alpha_l}{(l+1)}\Delta T^l \qquad (4)$$

# Heat diffusion 2D Model

To model the temperature diffusion in simple geometry sample, a 2D heat propagation waves simulation is conducted using Fourier equations:

$$\frac{\partial u}{\partial t} = \alpha \left( \frac{\partial^2 u}{\partial x^2} + \frac{\partial^2 u}{\partial y^2} \right) \qquad (5)$$

where $u(x, y, t)$ is the temperature at position $(x, y)$ and time $t$. $\alpha$ is the thermal diffusivity, defined as $\alpha = \rho c_p k$, where $k$, $\rho$, and $c_p$ are the material's thermal conductivity, density, and specific heat capacity, respectively. Equation (5) was solved numerically using a finite difference method (Forward Time Centered Space method) with Dirichlet boundary conditions. The initial temperature was set to 300 K across the film. To ensure numerical stability, the time and space steps were chosen based on the Courant–Friedrichs–Lewy (CFL) condition.

For the simulation, we used a 2D aluminum film with parameters $k = 237$ W.m$^{-1}$.K$^{-1}$, $\rho = 2700$ kg.m$^{-3}$ and $c_p = 897$ J.kg$^{-1}$.K$^{-1}$.[6] The temperature at the simulation edges was fixed at 300 K. The heat source, modeled to approximate the pulsed laser, was a circular Gaussian with a full width at half maximum (FWHM) of ~2 µm, providing energy for 35 ns. The heat source temporal profile is modeled on the PINEM signal temporal profile using an asymmetric trapezoidal function (**Figure S1a**). The energy deposited was treated as the only free parameter to fit the experimental data. Heat convection was neglected due to the low pressure in the sample chamber (~10$^{-9}$ mbar), and heat radiation effects were found to be negligible for the aluminum film.[7]

# Data treatment

For the three cases presented in this study, the different EELS signals were fitted in time to follow either their intensities (phonons) or positions (exciton and plasmons). This is realized using the Hyperspy library, which provides different modeling functions depending on the spectral peak of interest.[8] This section details the data fitting procedures used in each case. For every measurement, the time resolution is given by the highest frequency clock of the Timepix of about 0.2604 ns. However, each signal was temporally binned in order to have enough signal. Moreover, PCA was applied with different components threshold for the reconstruction.

### Phonons in silicon nitride membrane

We use a temporal binning factor coefficient of 20, for a final temporal resolution of approximately 5 ns. 20 components are used to reconstruct the data from PCA (see **Figure S3a** for the comparison before and after PCA application). The two gain and loss energy phonon peaks are fitted with a Gaussian function, after having determined the ZLP background shape with a Voigt function (**Figure S3b**). Note that the energy of the phonon peak (ΔE in **Eq. (1)**) was considered constant for the calculation of temperature profile, due to the low energy resolution of the current experiment.

The uncertainty related on temperature measurement is obtained by uncertainty propagation calculation. If we note $I_g$ and $I_l$ the gain and loss intensities, respectively, related to the phonon peak at energy $E_{ph}$, we can obtain from **Eq. (1)**:

$$T = \frac{E_{ph}}{K_b \ln(I_l/I_g)} \quad \rightarrow \quad \sigma_T = \frac{T}{\ln(I_l/I_g)} \sqrt{\left(\frac{\sigma_{I_l}}{I_l}\right)^2 + \left(\frac{\sigma_{I_g}}{I_g}\right)^2} \qquad (6)$$

where $\sigma_{I_g}$ and $\sigma_{I_l}$ represent the standard deviations of the gain and loss phonon peak intensities, respectively. For both gain and loss peaks, the net intensity is determined by subtracting the background contribution $I_i = I_{tot} - I_i^B$ where $I_i$ represents the net peak intensity (either loss or gain), and $I_i^B$ denotes the background intensity. Since both $I_i$ and $I_i^B$ are subject to Poissonian statistics, the uncertainty in $I_i$ must account for the contributions of both terms. Consequently, the standard deviation of $I_i$ is given by:

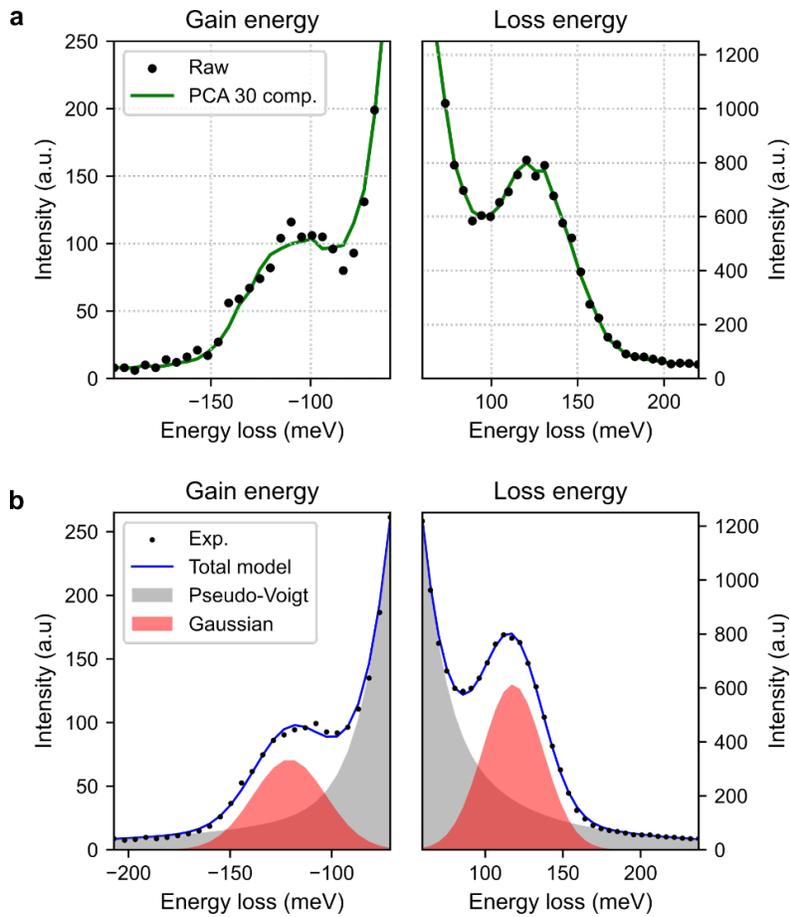

**Figure S2. Data treatment applied to silicon nitride phonons spectra.** (a) Effect of PCA reconstruction data noise. (b) Model fitting of the gain and loss energy parts of the spontaneous phonons excitation.

$$\sigma_{I_i} = \sqrt{I_i + 2I_i^B} \qquad (7)$$

$I^B$ is calculated by integrated the extended Split Voigt function used as a background model in an energy window corresponding to the Full Width at Tenth Maximum (FWTM) of both plasmon peaks.

**Plasmons in Al thin film**

To measure the temperature in a thin film aluminum, we need to measure the displacement of the bulk plasmon as a function of time and/or spatial position. To do so, a 2 components model is used, including (i) a 1$^{st}$ order polynomial general offset and (ii) a volume Drude plasmon function defined as:[4]

$$f(E) = I_0 \frac{E(\Delta E_p)E_p^2}{(E^2 - E_p^2)^2 + E(\Delta E_p)^2} \qquad (8)$$

where $I_0$, $E_p$ and $\Delta E_p$ are the peak intensity, the plasmon peak energy and the plasmon width at half maximum (FWHM), respectively. Before fitting, a 3 components PCA reconstruction is applied. **Figure S4** presents the results of the plasmon peak fitting.

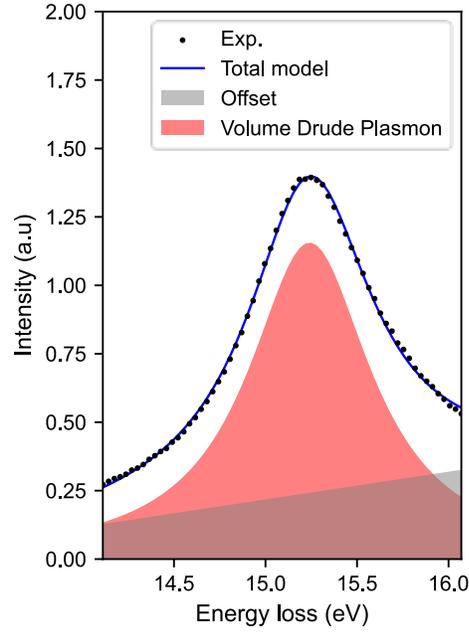

**Figure S3. Aluminum plasmon peak fit model**. The gray and red curves detail the 2 components used in the model.

The uncertainty propagation calculation from **Eq. (4)** leads to the following temperature standard deviation:

$$\sigma_T = \frac{1}{E_p(T_0)\sqrt{\Delta}}\,\sigma_{E_p} \qquad \text{where} \qquad \Delta = 3\left[a_1\left(1 - \frac{E_p}{E_p(T_0)}\right) + \frac{3}{4}a_0^{\,2}\right]$$

where $\alpha_l$ represents the $l^{\text{th}}$ parameter that describe the dependence of thermal expansion on temperature (measured by Chmielewski *et al.*[5]), and $\sigma_{E_p}$ is the standard deviation associated with the plasmon peak energy, determined from the fit quality.

**Excitons in TMDs**

To describe the EELS optical range spectrum obtained from a WS$_2$ (and measure the evolution on the peaks in time), a model consisting of 5 different components is build. After applying a 100x temporal binning factor, the time-resolved spectra are fitted with the following model (**Figure S5**):

- A pure offset, accounting for the usual background obtained in this region from e.g. Cherenkov radiation;
- An exponential decay to fit the ZLP tail;
- 2 gaussians for the A and B excitons peaks
- A mixed of 3 gaussians for the C exciton.

The material temperature evolution is tracked using the position of A exciton component.

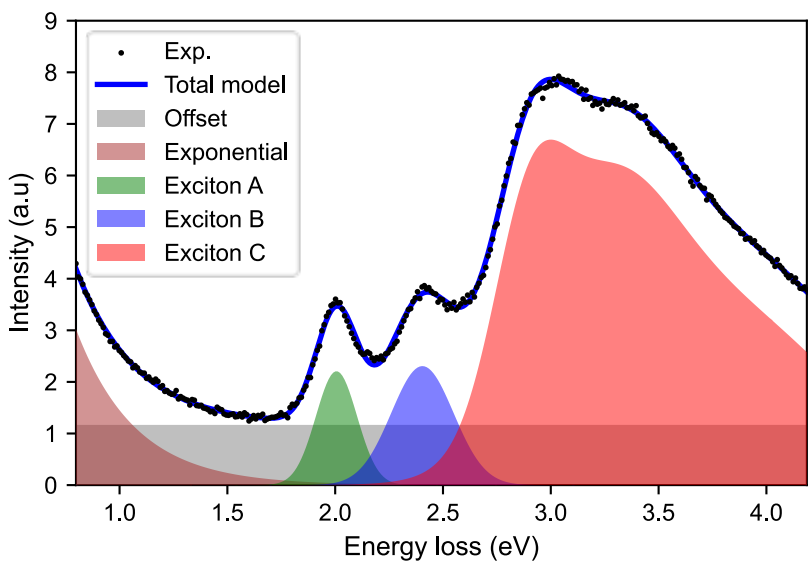

**Figure S4. Excitons peaks fit in WS$_2$.** Each component is separately represented.

# References

(1) Auad, Y.; Baaboura, J.; Blazit, J.-D.; Tencé, M.; Stéphan, O.; Kociak, M.; Tizei, L. H. G. Time Calibration Studies for the Timepix3 Hybrid Pixel Detector in Electron Microscopy. *Ultramicroscopy* **2024**, *257*, 113889. https://doi.org/10.1016/j.ultramic.2023.113889.

(2) La Grange, T.; Cattaneo, P.; Carbone, F.; Weaver, B.; Sapozhnik, A.; Yang, Y.; Raja, A. S.; Kippenberg, T. J. Electron Spectrometer Calibration Method. EP4290549A1, December 13, 2023.

(3) Tizei, L. H. G.; Auad, Y.; Blazit, J.-D.; Tencé, M.; Kociak, M. Method and System for Calibrating a Charged-Particle Spectrometer. WO2024046668A1, March 7, 2024.

(4) Egerton, R. F. *Electron Energy-Loss Spectroscopy in the Electron Microscope*; Springer US: Boston, MA, 2011. https://doi.org/10.1007/978-1-4419-9583-4.

(5) Chmielewski, A.; Ricolleau, C.; Alloyeau, D.; Wang, G.; Nelayah, J. Nanoscale Temperature Measurement during Temperature Controlled in Situ TEM Using Al Plasmon Nanothermometry. *Ultramicroscopy* **2020**, *209*, 112881. https://doi.org/10.1016/j.ultramic.2019.112881.

(6) Zhang, A.; Li, Y. Thermal Conductivity of Aluminum Alloys—A Review. *Materials* **2023**, *16* (8), 2972. https://doi.org/10.3390/ma16082972.

(7) Mecklenburg, M.; Hubbard, W. A.; White, E. R.; Dhall, R.; Cronin, S. B.; Aloni, S.; Regan, B. C. Nanoscale Temperature Mapping in Operating Microelectronic Devices. *Science* **2015**, *347* (6222), 629–632. https://doi.org/10.1126/science.aaa2433.

(8) De La Peña, F.; Prestat, E.; Fauske, V. T.; Burdet, P.; Lähnemann, J.; Jokubauskas, P.; Furnival, T.; Nord, M.; Ostasevicius, T.; MacArthur, K. E.; Johnstone, D. N.; Sarahan, M.; Taillon, J.; Aarholt, T.; Pquinn-Dls; Migunov, V.; Eljarrat, A.; Caron, J.; Francis, C.; T.

Nemoto; Poon, T.; Mazzucco, S.; Actions-User; Tappy, N.; Cautaerts, N.; Suhas Somnath; Slater, T.; Walls, M.; Winkler, F.; Ånes, H. W. Hyperspy/Hyperspy: Release v1.7.2, 2022. https://doi.org/10.5281/ZENODO.7090040.